Simple determination of the orbit of a comet, when it is possible to observe its passage across the ecliptic twice

E547 -- *Determinatio facilis orbitae cometae, cuius transitum per eclipticam bis observare licuit*

Originally published in *Acta Academiae Scientarum Imperialis Petropolitinae pro Anno MDCCLXXX*, 1783, pp. 243 - 254.
According to the records, it was presented to the Petersburg Academy on August 19, 1776.

Translated and Annotated[1]
by
Sylvio R. Bistafa[*]
June 2020

Foreword

This is the translation from Latin of one of the Euler's papers on celestial mechanics, which considers the determination of a comet's parabolic orbit, with the Sun at the focus, from two astronomical observations from the earth, when the comet crosses the ecliptic at the ascending and descending nodes. The key point of the calculation is the solution of a fourth degree polynomial, from which the determination of the orbital parameters are determined from one of its roots.

Author
L. Euler

§1

But if both transits of the ecliptic plane by a certain Comet may be observed, although this is very rare, nonetheless, this case is very fitting to be considered with maximum attention, because by this direct method, two such observations suffice, for perfectly determining the parabolic orbit of a comet, which does not have to recourse to any approximations. On the other hand, if from other observations, the motion of comets can not at all be determined by a direct method, and after many attempts trying to dig it out by approximations, and eventually unable to succeed, be then such direct method disclosed, by whose power the orbits of Comets may certainly be determined from these observations.

§. 2. Let us then consider that such a Comet becomes visible, whose passage across the ecliptic it is possible to observe twice, so that the latitude in both observations has been revealed to be zero. For these times, it is necessary that the nodal line itself be traversed by the Comet, certainly once in the ascending node, and the other truly in the descending [node], then, we would recall the two observations for the calculation in the following way. Fig. 1 represents the plane of the ecliptic, in which the point $S$ is the center of the Sun, and at the time of the prior observation, the Earth will be at point $T$. Let the distance to the Sun

---

[1] The translator used the best of his abilities and knowledge to make this translation technically and grammatically as sound as possible. Nonetheless, interested readers are encouraged to submit suggestions for corrections as they see fitting.
[*] Corresponding address: sbistafa@usp.br

be $ST = a$. Meanwhile the comet will appear in the direction of $TZ$, and the angle $STZ$ will be known, which we will call $\alpha$. Since the comet will have zero latitude it will stick somewhere on the line $TZ$ itself, whose position let us take to be point $Z$, so that the line drawn from the Sun, $SZ☊$ will be the nodal line, or the intersection of the orbit of the Comet with the ecliptic, which position is still unknown. Let us call the angle $TS☊ = \phi$, which so far is the only unknown that is necessary to be introduced into the calculation. Thus, the external angle will be given by $TZ☊ = \alpha + \phi$, whence, since the distance $ST = a$ is known, the distance of the Comet to the Sun will be given by $SZ = \frac{a \sin \alpha}{\sin(\alpha + \phi)}$.

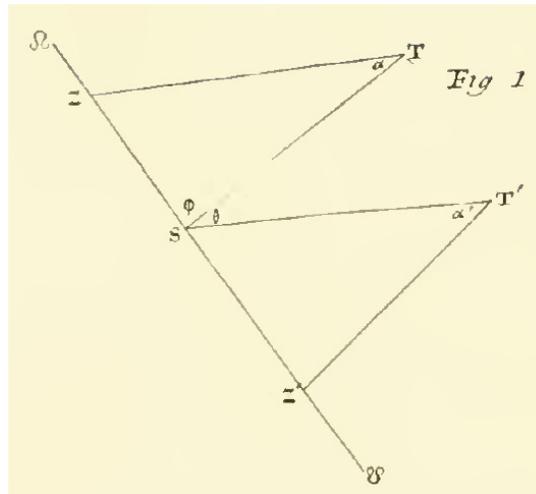

§. 3. After the time $= \theta$ has elapsed, which we express by the mean motion of the Sun, so that $\theta$ designates a certain angle, whose measure will be the arc $\theta$ itself in a circle which radius $= 1$, while the whole periphery $2\pi$ of this circle represents the extent of one year, when the comet in the actual ecliptic is observed without latitude a second time; meanwhile the Earth advanced through the angle $TST' = \theta$, and the comet will appear in the direction $T'Z'$, creating the angle $ST'Z' = \alpha'$. Then it is necessary that at this time the comet had traversed backwards in the nodal line $S\Omega$ continued to $S☋$ and approaching the point $Z'$ itself. Therefore, since the angle $T'SZ' = 180° - \phi - \theta$, then the external angle $T'Z'☋ = 180° + \alpha' - \phi - \theta$, and then the angle $T'Z'S = \phi + \theta - \alpha'$. Setting the distance $ST' = a'$, then

$$\sin(\phi + \theta - \alpha') : a' = \sin \alpha' : SZ',$$

whence, the distance $SZ' = \frac{a' \sin \alpha'}{\sin(\phi + \theta - \alpha')}$. Thus, we have defined two distances from the Comet to the Sun, namely

$$SZ = \frac{a \sin \alpha}{\sin(\alpha + \phi)} \quad \text{and} \quad SZ' = \frac{a' \sin \alpha'}{\sin(\phi + \theta - \alpha')}$$

which are exactly opposite to each other, such that the true anomaly described between them is $180°$.

§. 4. Figure 2 refers now to the actual plane of the orbit of the comet, in which $S$ is the center of the Sun and the line segment $☊☋$ is the intersection of the orbit of the comet with the plane of the ecliptic, in which points $Z$ and $Z'$ are the two observed locations of the Comet. Here, let us put the distances $SZ = f$ and $SZ' = g$, such that we have

$$f = \frac{a \sin \alpha}{\sin(\alpha + \phi)} \quad \text{and} \quad g = \frac{a' \sin \alpha'}{\sin(\phi + \theta - \alpha')},$$

in which the only unknown quantity that exists is, of course, the angle $\phi$. Moreover, certainly we know that the comet came from the location $Z$ to arrive at the location $Z'$ during the time $= \theta$. Then parabola $Z\Pi Z'$ would be the orbit of the comet described in the same time, whose axis is the line segment $\Pi S$, such that $\Pi$ is its perihelion, whose distance to the Sun is given as $S\Pi = p$,[2] and thus the focal parameter[3] of the orbit $= 2p$; then let us call the angle $\Pi SZ = \psi$, such that the angle $\Pi SZ' = 180° - \psi$. Hence, from the very nature of the parabola we have that

$$SZ = f = \frac{2p}{1 + \cos\psi} = \frac{p}{\cos^2\frac{\psi}{2}}$$

$$SZ' = g = \frac{2p}{1 - \cos\psi} = \frac{p}{\sin^2\frac{\psi}{2}}$$

from these equations the distance $p$ and the angle $\psi$ will be easily determined, if $f$ and $g$ are considered known quantities.

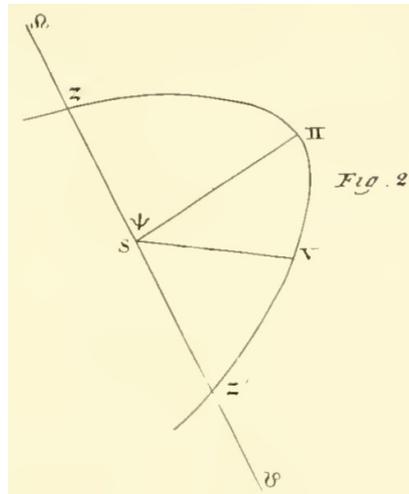

Fig. 2

§. 5. Therefore, since from the first equation $\cos^2\frac{\psi}{2} = \frac{p}{f}$, and from the second $\sin^2\frac{\psi}{2} = \frac{p}{g}$, then adding these $\frac{p}{f} + \frac{p}{g} = 1$, so $p = \frac{fg}{f+g}$; this value having been found $\cos^2\frac{\psi}{2} = \frac{g}{f+g}$ and $\sin^2\frac{\psi}{2} = \frac{f}{f+g}$; whence is given $\tan\frac{\psi}{2} = \sqrt{\frac{f}{g}}$ and $\sin\psi = \frac{2\sqrt{fg}}{f+g}$. Therefore, provided that both distances $f$ and $g$ have been obtained, the parabolic orbit of the Comet will be perfectly determined. In fact, both distances $f$ and $g$ and the angle $\phi$ are still unknown, so, in addition, we need another equation, which would allow these unknowns to be determined.

§. 6. However, this new equation, require to us the consideration of the time $\theta$, during which the Comet passes from the location $Z$, through $\Pi$, and reaches $Z'$, since, meanwhile, the area described by the Comet is proportional to this time. Then, if for this area is written $Z\Pi Z' = S$, and the mean distance of the Earth to the Sun is indicated by the letter $c$, and because the focal parameter of this orbit $= 2p$, then from

---

[2] Here $p$ represents the distance from the vertex to the focus of the parabola.
[3] The distance between the directrix and focus.

the theory of the motion of the planets we will have that $S = \frac{1}{2}\Theta c\sqrt{2cp}$;[4] and, on this account, it will be necessary that we only seek the amount of this area.

§. 7. Now that we have the angle $\Pi SZ = \psi$, let us meanwhile call the distance $SZ = z$, and because of the nature of the parabola, $z = \frac{p}{\cos^2\frac{\psi}{2}}$. Let us seek first the area $\Pi SZ$, which let us call $\Sigma$, so that $d\Sigma = \frac{1}{2}z^2 d\psi$, then

$$d\Sigma = \frac{\frac{1}{2}p^2 d\psi}{\cos^4\frac{\psi}{2}}.$$

To integrate this formula, let us put $\tan\frac{\psi}{2} = t$, and then

$$\sin\frac{\psi}{2} = \frac{t}{\sqrt{1+t^2}} \quad \text{and} \quad \cos\frac{\psi}{2} = \frac{1}{\sqrt{1+t^2}},$$

and

$$\frac{1}{2}d\psi = \frac{dt}{1+t^2},$$

Therefore,

$$d\Sigma = p^2 \frac{dt}{(1+t^2)\cos^4\frac{\psi}{2}} = p^2 dt(1+t^2),$$

whence, upon integration, we find the area $\Sigma = p^2 t + \frac{1}{3}p^2 t^3$, and, of course, $t = \tan\frac{\psi}{2}$.

§. 8. Hence, also, the other area $\Pi SZ'$ is easily derived, which we call $\Sigma'$, if in place of $\psi$ in the preceding formula we write $180° - \psi$, and then writing $90° - \frac{\psi}{2}$ in place of $\frac{\psi}{2}$, and since its tangent is $\cot\frac{\psi}{2} = \frac{1}{t}$, immediately is found the area $\Sigma' = \frac{p^2}{t} + \frac{p^2}{3t^3}$. With these two areas joined together, the total area would amount to

$$Z\Pi Z' = S = \Sigma + \Sigma' = \frac{p^2}{3t^3}(t^2+1)^3.$$

Hence, since $S = \frac{p^2}{3}\left(\frac{t^2+1}{t}\right)^3$, and by introducing the angle $\psi$ again, then

$$S = \frac{p^2}{3}\left(\frac{1}{\sin\frac{\psi}{2}\cos\frac{\psi}{2}}\right)^3, \text{ or}$$

$$S = \frac{p^2}{3}\left(\frac{2}{\sin\psi}\right)^3 = \frac{8p^2}{3\sin^3\psi}.$$

§. 9. Therefore, the area $S$ just found and the relation mentioned above between the time and this area, give us this equation[5]

---

[4] The area of the ellipse $A = \pi c\sqrt{2cp}$, where $2p$ is the so called semi-latus rectum of the ellipse. Considering Kepler's 2nd law, which states that an imaginary line joining a planet and the sun sweeps out an equal area of space in equal amounts of time, then the area $S$ swept during the time $\Theta$, is such that $\frac{S}{A} = \frac{\Theta}{2\pi}$ is constant, then $S = \frac{1}{2}\Theta c\sqrt{2cp}$.

$$\frac{8p^2}{3\sin^3\psi} = \frac{1}{2}\Theta c\sqrt{2cp},$$

which, once divided by $\sqrt{p}$ gives

$$\frac{8p\sqrt{p}}{3\sin^3\psi} = \frac{1}{2}\Theta c\sqrt{2c},$$

or

$$\frac{8p\sqrt{p}}{\sin^3\psi} = \frac{3}{2}\Theta c\sqrt{2c},$$

whence, by extracting the cubic root, gives

$$\frac{2\sqrt{p}}{\sin\psi} = \sqrt[3]{\frac{3}{2}\Theta c\sqrt{2c}}.$$

However, we see above that

$$p = \frac{fg}{f+g} \quad \text{and} \quad \sin\psi = \frac{2\sqrt{fg}}{f+g}, \text{ whence, } \frac{2\sqrt{p}}{\sin\psi} = \sqrt{f+g},$$

so that we are led to this equation

$$\sqrt{f+g} = \sqrt[3]{\frac{3}{2}\Theta c\sqrt{2c}}$$

which once squared gives $f + g = c\sqrt[3]{\frac{9}{2}\Theta^2}$. Next, we substitute the values for $f$ and $g$ found above, and then we have the following equation

$$\frac{a\sin\alpha}{\sin(\alpha+\phi)} + \frac{a'\sin\alpha'}{\sin(\phi+\theta-\alpha')} = c\sqrt[3]{\frac{9}{2}\Theta^2}$$

in which there remains only one unknown, namely the angle $\phi$, and the whole matter is conveyed in such a way that the value of the angle $\phi$ is obtained from this equation. In what follows, which we shall show how this can be most conveniently accomplished.

§. 10. The solution of this equation can be more easily accomplished when we represent it in this form

$$\frac{A}{\sin(\omega-\gamma)} + \frac{B}{\sin(\omega+\gamma)} = C,$$

so that

$$A = a\sin\alpha, B = a'\sin\alpha', C = c\sqrt[3]{\frac{9}{2}\Theta^2};$$

and then, in fact,

---

[5] Euler gives no justification for the equality between the area described by the Earth and that described by the Comet.

$$\omega - \gamma = \alpha + \phi \quad \text{and} \quad \omega + \gamma = \phi + \theta - \alpha'$$

henceforth

$$2\omega = \alpha + 2\phi + \theta - \alpha' \quad \text{and} \quad 2\gamma = \theta - \alpha - \alpha'$$

and, therefore, the last angle is known right away since $\gamma = \frac{\theta - \alpha - \alpha'}{2}$; while, truly, the unknown angle will be $\omega = \frac{\alpha + 2\phi + \theta - \alpha'}{2}$, such that if the angle $\omega$ is determined, then the desired angle will be $\phi = \omega + \frac{\alpha' - \alpha - \theta}{2}$. Now, to find the angle $\omega$, our equation freed from fractions will be

$$A \sin(\omega + \gamma) + B \sin(\omega - \gamma) = C \sin(\omega + \gamma) \sin(\omega - \gamma)$$

which unfolded further, gives the following

$$A \sin\omega \cos\gamma + A\cos\omega \sin\gamma + B \sin\omega \cos\gamma - B\cos\omega \sin\gamma = C[\sin^2\omega \cos^2\gamma - \cos^2\omega \sin^2\gamma]$$

which the angle $\omega$ may be determined. Then, finally, if we put $\sin\omega = s$, then $\cos\omega = \sqrt{1 - s^2}$; and for this equation to be liberated from irrationality, it would have to be squared again. However, this we can now avoid with the following operation.

§. 11. Of course, this equation can be at once reduced to a rational expression by considering $\tan\frac{\omega}{2} = x$, and then $\sin\frac{\omega}{2} = \frac{x}{\sqrt{1+x^2}}$ and $\cos\frac{\omega}{2} = \frac{1}{\sqrt{1+x^2}}$, whence, further

$$\sin\omega = \frac{2x}{1 + x^2} \quad \text{and} \quad \cos\omega = \frac{1 - x^2}{1 + x^2}$$

with these values substituted, our equation will be

$$\frac{2x \cos\gamma}{1 + x^2}(A + B) + \frac{(1 - x^2) \sin\gamma}{1 + x^2}(A - B) = C\frac{(4x^2 \cos^2\gamma - (1 - x^2)^2 \sin^2\gamma)}{(1 + x^2)^2}.$$

Furthermore, for the sake of brevity let us put $(A + B) \cos\gamma = F$ and $(A - B) \sin\gamma = G$, and multiplying by $(1 + x^2)^2$, then our equation will be reduced to

$$2Fx(1 + x^2) + G(1 - x^4) = C(4x^2 \cos^2\gamma - (1 - x^2)^2 \sin^2\gamma),$$

which once expanded is reduced to the following equation of fourth degree:

$$(G - C \sin^2\gamma)x^4 - 2Fx^3 + 2Cx^2(1 + \cos^2\gamma) - 2Fx - C \sin^2\gamma - G = 0,$$

from which, it has two, or even all, of its roots real, they will give rise to the same number of solutions to our problem, of which, those that, in fact, should be considered can be easily defined, if any additional third observation is invoked in relief, from which the inclination of the orbit in relation to the ecliptic will become known.

§. 12. Therefore, from any root $x$ of the equation that was found, the angle $\omega$ is immediately obtained, and since $\tan\frac{\omega}{2} = x$, then, in fact, from the knowledge of the angle $\omega$, this angle is derived

$$\phi = \omega + \frac{\alpha' - \alpha - \theta}{2}$$

from which the true position of the nodal line is known, since $\mathcal{S}ST = \phi$. Thereafter, in fact, from the knowledge of the angle $\phi$, also the distance of the Comet to the Sun will be defined, since

$$SZ = f = \frac{a\sin\alpha}{\sin(\alpha + \phi)} \quad \text{and} \quad SZ' = g = \frac{a' \sin\alpha'}{\sin(\phi + \theta - \alpha')},$$

but since,

$$f = \frac{a\sin\alpha}{\sin(\alpha - \gamma)} \quad \text{and} \quad g = \frac{a'\sin\alpha'}{\sin(\alpha + \gamma)},$$

knowing that $\gamma = \frac{\theta - \alpha - \alpha'}{2}$; then, from these discoveries, the distance of the perihelion to the Sun is obtained from

$$S\Pi = p = \frac{fg}{f + g},$$

and the angle $\Pi SZ = \psi$, will be known from $\tan\frac{\psi}{2} = \sqrt{\frac{f}{g}}$. Therefore, in this way, the whole orbit of the Comet will be determined exactly, and nothing else remains, provided its inclination in relation to the ecliptic becomes known, which will be easily accomplished from any other observation where the latitude of the Comet is observed.

§. 13. Also from these, which have just already considered, the time, in which the Comet had traversed through the perihelion $\Pi$, will be possible to be determined with little difficulty. If for this time, during which the comet from the $Z$ location had arrived at the perihelion we put $= T$, for which we found above the area $\Pi SZ = \Sigma = p^2\left(t + \frac{1}{3}t^3\right)$, and since $t = \tan\frac{\psi}{2}$, we will have from the relation between area and time $\Sigma = \frac{1}{2}Tc\sqrt{2cp}$, whence, the desired time will be

$$T = \frac{2\Sigma}{c\sqrt{2cp}} = \frac{2p\sqrt{p}\left(t + \frac{1}{3}t^3\right)}{c\sqrt{2c}}.$$

Therefore, since $p = \frac{fg}{f+g}$ and $t = \sqrt{\frac{f}{g}}$, then, with these values been substituted, the desired time will be found as

$$T = \frac{2f^2\left(g + \frac{1}{3}f\right)}{(f + g)^{3/2}c\sqrt{2c}}$$

or, since $f + g = c\sqrt[3]{\frac{9}{2}\Theta^2}$, and thus

$$(f + g)^{3/2} = c\sqrt{c}\frac{3\Theta}{\sqrt{2}} \quad \text{and} \quad T = \frac{2f^2\left(g + \frac{1}{3}f\right)}{3c^3\Theta},$$

in this way, the transit time through the perihelion will be known, for a regular root $x$ taken from that fourth degree equation, the one which other subsidiary observations called upon will be settled soon.

§. 14. Most remarkable is the equation to which we are led, which is $f + g = c\sqrt[3]{\frac{9}{2}\Theta^2}$, since $f + g$ express the distance between the points $Z$ and $Z'$, that is, [the distance between] two locations of the comet in its orbit, which seen from the Sun, are themselves in opposition, through which, it is possible to generate the location of the nodal line by means of any segment drawn through the Sun, and from only one time, during which the Comet is brought from one extremity of this segment to the next, the extent of this segment, or the distance between these two locations opposing each other can be ascribed, being the requirements of this operation included in the following theorem.

## A General Theorem for the Motion of Comets in Parabolic Orbits

§. 15. If a Comet is moved around the Sun on any parabolic orbit $F\Pi G$, with its focus located at $S$, and knowing the time during which from any location $F$ it is taken to its opposite $G$, from these, it is possible to fully assign the extent of this segment $FG$. So, from solar tables, is that time which corresponds to the mean motion of the Sun, which is given in signes[6], degrees and minutes, and written as $= N$ degrees; then, it happens that $360°:N =$ the whole perimeter of the circle $2\pi: \Theta$, and then $\Theta = \frac{2\pi N°}{360°}$. Furthermore, if the mean distance of the Sun to the Earth is called $= c$, the distance of those two locations $F$ and $G$, or the segment $FG$ will always be equated to this formula: $c \sqrt[3]{\frac{9}{2}\Theta^2}$. From which, it is recognized that the cube of the segment $FG$ is always proportional to the square of the time during which the Comet will reach from $F$ to $G$.

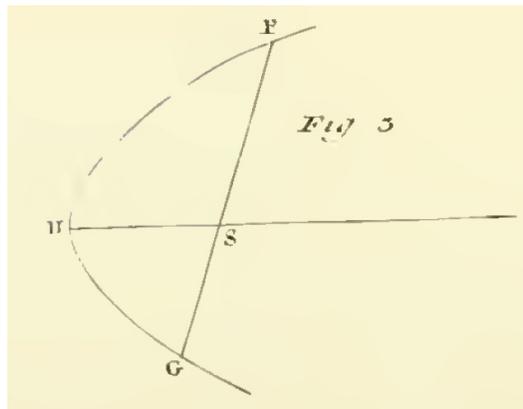

Fig. 3

§. 16. However, since the last equation of fourth degree that we have found has many roots $x$, let us see if we can discern which one from these, in fact, gives the correct locus. To this the end, when we call for the help of a certain third observation, in which both the longitude and the latitude had been determined, and also, for the time of the orbit already known, will be sought the position of the comet in its orbit, that is $V$, so that, for this time, the distance of the Comet to the Sun will be known, and the angle, that is, the argument of the latitude $\mathcal{S}SV$. Set the distance $SV = v$ and the angle $\mathcal{S}SV = \eta$.

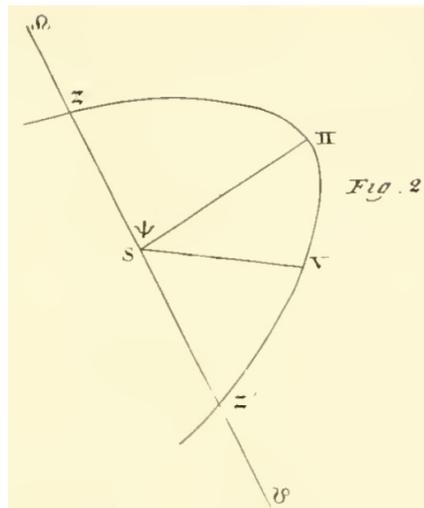

Fig. 2

---

[6] 1 Signe (1ˢ) => 30°.

§. 17. Now, in the plane of the ecliptic again, let the segment $S\Omega$ be the nodal line of the Comet, and at the time of the third observation, the Earth will be at $T$, so that the distance $ST = b$. Now let $Z$ be the location of the comet, from which let the perpendicular $ZX$ be dropped in the direction of the plane of the ecliptic and let the segments $TXU$ and $TZ$ be drawn. Since we assumed the latitude and the longitude have been observed, the angles $STU$ and $TUZ$ will be known. Once the drawn segment $SZ$ is known, then the angle $\Omega Sz = \eta$ will be known, together with the actual distance $SZ = v$. Then, if from the point $X$, the normal line $XP$ to the nodal line is drawn, and the segment $PZ$ is drawn, then $PZ = v \sin \eta$ and $SP = v \cos \eta$. Hence, since the point $X$ is known, so is the segment $TX$, and further the perpendicular $ZX$, from which is determined the inclination of the comet's orbit, that is the angle $XPZ$. Likewise, it is also possible, if the calculation is set up this way, not only to be able to find the inclination of the orbit to the ecliptic, but also it will be easy to judge whether other conclusions made from the theory, namely the distance $v$ and the angle $\eta$ correspond to the given observations or not.

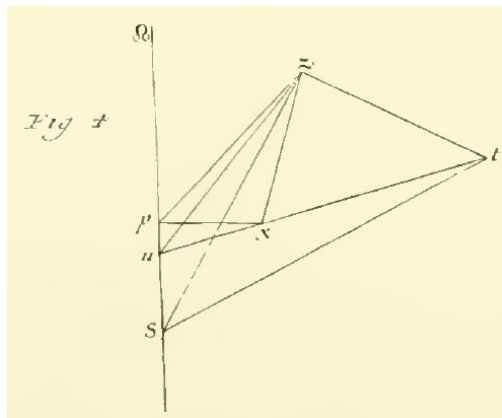

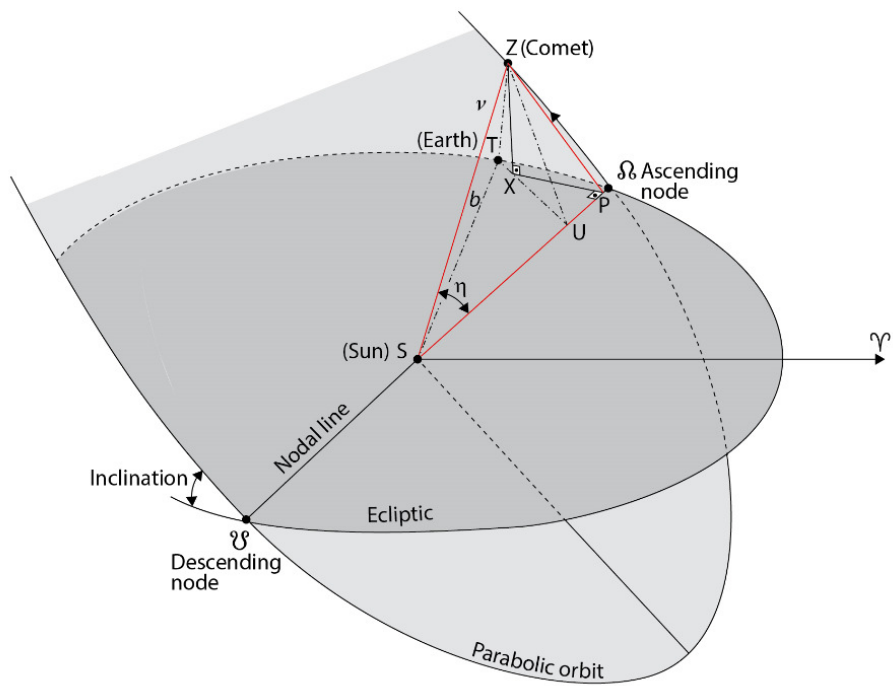

---

[7] This figure was added by the Translator.

§. 18. Let this becomes even clearer, once the next steps are followed. 1) From the triangle $STU$, in which all the angles along with the side $ST$ are given, then sides $SU$ and $TU$ will be known. 2) From the distance $SZ$ and with the angle $☊SZ$, the segments $SP$ and $PZ$ will be given, evidently, after the perpendicular $ZP$ has been drawn from the location of the comet $Z$ to the nodal line. 3) Next, therefore, the interval $UP$ will become apparent, and thus, the normal $PX$, drawn in the plane of the ecliptic, from the point $P$ to the nodal line, will intercept the line $TU$ at point $X$, and from the interval $UP$ and the angle $TUP$, the interval $UP$ as well as $UX$ are defined, which having been laid out on $TU$ the interval $TX$ will remain. 4) Further, from the observed latitude or angle $XTZ$, the elevation of the comet above the ecliptic is defined, obviously the segment $XZ$ itself, and the three sides of the right triangle $ZPX$ are defined, therefore, in fact, $ZP^2 = PX^2 + XZ^2$, observing that this is the exact root of that fourth degree equation that was assumed; however, in case this equality does not hold here, another root of that equation should be considered, and another calculation will have to be established in a similar way. Then, after the true root has been found, such that $PZ^2 = PX^2 + XZ^2$, then, the angle $ZPX$ will give the inclination of the orbit of the comet in relation to the ecliptic. Therefore, in this way, in cases in which it is possible to observe both passages of a comet through the ecliptic, it will be possible to determine its true parabolic orbit, absolutely, without any trial or approximation.

_________________